\documentclass{elsart}

\usepackage{graphics}
\usepackage{natbib}
\usepackage{amssymb}

\begin{document}

\begin{frontmatter}

\title{The origin of the mid- and far-infrared emission in AGN}

\author[eso,iac]{M. Almudena Prieto}
%\ead{aprieto@eso.org}
\author[iac]{Ana P\'erez-Garc\'{\i}a}
%\ead{apg@ll.iac.es}
\author[iac]{Jos\'e M. Rodr\'{\i}guez-Espinosa}
%\ead{jre@ll.iac.es}

\address[eso]{European\,Southern\,Observatory, Karl-Schwarzschild-Strasse\,2, D--85748~Garching}
\address[iac]{Instituto de Astrof\'{\i}sica de Canarias,
E-38200~La~Laguna, Tenerife} 

\begin{abstract}
The relation between the X-ray emission, the infrared emission and 
the coronal line emission in a sample of the 
brightest known Seyfert galaxies  is analyzed. 
A close relationship between the  absorption-corrected soft X-ray emission 
 and both the mid-infrared and the coronal line emission is found among the 
 Seyfert type\,2 objects studied in this work. 
 Being both the coronal line  and the X-ray emissions main
tracers of the AGN activity, their relation  with the mid-infrared
emission adds further support to the claim   that the  mid-infrared
emission is primarily due to dust heated
directly by the central AGN. No apparent relationship
 between the soft X-ray emission
and  the coronal line and/or the mid-infrared emissions is seen within 
Seyfert\,1 objects,   the number of objects of this class in the sample 
is comparatively smaller though.
That discrepancy is partially due to the fact that the soft X-ray emission in 
the Seyfert\,1s is 
systematically larger by at least one order of magnitude
than that in  their Seyfert\,2 counterparts. 
Finally, the  hard X-ray emission in both Seyfert types 
shows unrelated to the mid-infrared and coronal line emissions.
\end{abstract}

\end{frontmatter}

\vspace{-12mm}
\section{Introduction}
\vspace{-6mm}
Coronal lines arising in the spectra of AGNs are unique tracers of the
pure AGN power mechanism. These lines require ionization potentials
(IP) beyond 50\,eV, and thus their study provides clues on the ionizing
continuum spectrum, which on the other hand is difficult to probe
from observations, because of the heavy absorption.

In a previous work (Prieto et al. 2000) the emission from the two
 strongest coronal lines, [O\,IV]$\lambda$25.9\,$\mu$m and 
[Ne\,V]$\lambda$14.5\,$\mu$m, in the ISO
 spectra of a sample of bright Seyfert galaxies was found to be
 related to the mid-infrared continuum emission arising in these
 objects. Because of the nuclear origin of coronal lines, this
 relation was interpreted as the mid-infrared being due to
 emission by dust heated primarily by the central AGN.

In this paper, that relation is explored further on the basis of X-ray
data available for the galaxies in the sample. The reasoning behind is
as follows. Being the IP of $O^{2+}$ about 50\,eV and that of
$Ne^{3+}$ 100\,eV, a correlation between the [O\,IV] and [Ne\,V] line
fluxes and the soft X-ray emission might be anticipated. If, as argued
in Prieto et al. (2000), the emission in the mid-infrared is due to
dust mostly  
heated within the AGN nuclear region, a further relation between the
mid-infrared continuum emission and the soft X-ray emission is also
expected.  We explore those predictions, taking as a reference a sample
of the brightest known Seyfert galaxies which are studied by us in
detail with ISO. This sample contains seven Seyfert type\,2 and four
Seyfert type\,1 and includes prototypes objects such as NGC\,1068, 
NGC\,4151
and Circinus. The ISO coronal line spectrum of the sample is studied
in Prieto and Viegas (2000); the mid- to far-infrared continuum emission is
studied in P\'erez-Garc\'{\i}a et al. (2000); the relationship between the infrared
continuum and the coronal line emission is discussed in Prieto et al.
(2000).

\vspace{-6mm}
\section{Observational data}
\vspace{-7mm}
The sample of Seyfert galaxies used in this work is originally
presented in Prieto \& Viegas (2000). All the galaxies were observed
with ISO SWS at the wavelengths of [O\,IV]$\lambda$25.9\,$\mu$m, 
[Ne\,V]$\lambda 14.3$\,$\mu$m, [Mg\,VIII]$\lambda 3.02$\,$\mu$m and 
[Si\,IX]$\lambda 2.58$\,$\mu$m. Integrated line fluxes from the two
strongest lines in the Seyfert spectra, [O\,IV] and [Ne\,V], are used
in this work. 
Continuum fluxes at 16, 25 and 60\,$\mu$m measured with ISOPHOT
 and 10\,$\mu$m ground-based data  
existing for all the sources are also used. Those data are taken from
P\'erez-Garc\'{\i}a et al. (2000).

\vspace{-1mm}
The soft X-ray emission, namely in the 0.2--2.4\,keV region, is primarily
derived  from the ROSAT PSPC. For homogeneity purposes, when available
PSPC pointing observations, the absorption-corrected X-ray emission is derived
by us after applying an absorbed power-law fit to the PSPC spectrum.
In all cases, statistically acceptable fits were derived, with all the
fitting parameters; column density, power-law spectral index and
normalization, derived  within reasonable constraints. For those
cases in which the derived column density was lower than the
corresponding Galactic value, this was used instead as a fixed parameter
in the fit and the integrated X-ray emission derived accordingly. In
either case, the
error associated  to the X-ray flux refers to 1\,$\sigma$ uncertainty in the
fit.  For those sources with known complex spectrum and/or morphology,
integrated absorption corrected fluxes were taken from the literature. In these
cases, the error considered represents the amplitude of the variation
between different measurements in the literature. 

Finally, for the hard X-ray emission, namely the 2--10\,keV region,
absorption-corrected fluxes are in all cases taken from the
literature.  Preference was given to ASCA or BESPOSAX data.  
In general, when several absorption-corrected fluxes were reported, an
average value is used and the associated error reflects the amplitude
of the variation.  The sample includes three known Compton thick
sources: NGC\,1068, Mrk\,533 and Circinus. In these cases, the associated
hard X-ray emission may be subjected to large uncertainties as the
derived values are strongly model dependent.
Further details on the analysis of the X-ray data are given in Prieto
et al. (2000).

\begin{figure}
%\vspace{22.0cm}
\begin{center}
\resizebox{14cm}{!}{\includegraphics{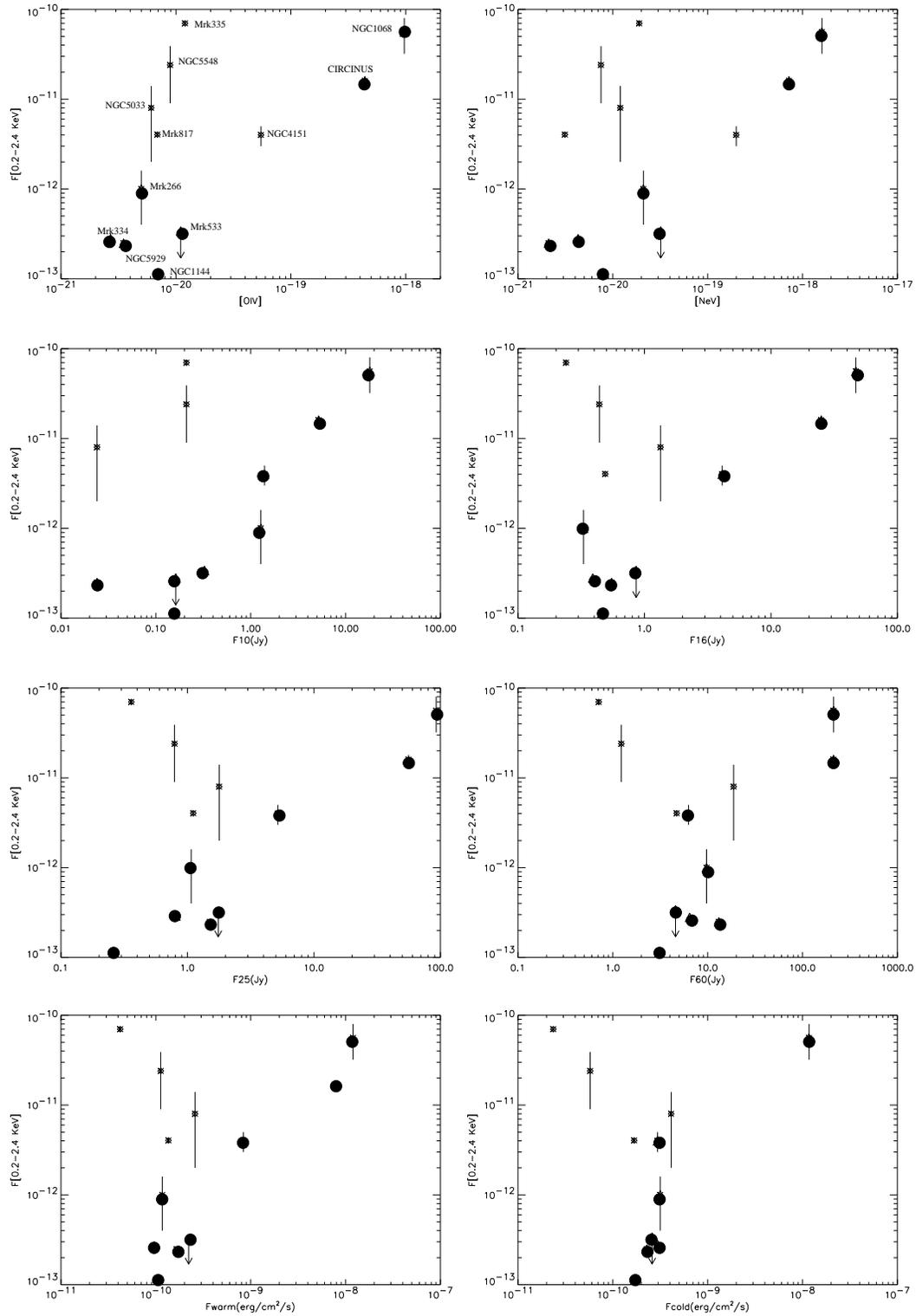}}
\parbox[b]{14cm}{
\vspace{2mm}
\caption[] {The relation between the soft X-ray emission and
the coronal line and mid-infrared emission in Seyfert galaxies. Filled
circles represent Seyfert 2 objects and  the intermediate
Seyfert type NGC4151.}
}
\label{Figure 1}
\end{center}
\end{figure}

\vspace{-8mm}
\section{Results}
\vspace{-6mm}
For all the galaxies in the sample, Fig.~1 compares the integrated
absorption-corrected soft X-ray emission with respectively the coronal
line fluxes of [O\,IV] and [Ne\,V] (first row), and with the 10, 16, 25
and 60\,$\mu$m continuum emission (second and third rows in Fig.~1).
The comparison with the respective warm and cold infrared emission as
defined in P\'erez-Garc\'{\i}a et al. (2000) is shown in the last row of the
figure.  Seyfert\,2 galaxies are marked with circles to differentiate
them from their Seyfert\,1 counterparts. 

A rather scatter diagram is present in all cases; yet, when the
comparison is restricted to the Seyfert\,2 objects, a trend is
apparent. This is particularly clear when considering the coronal line
emission and the 10\,$\mu$m flux emission; the trend becomes less
clear when involving colder infrared fluxes and almost vanishes when
considering the 60\,$\mu$m flux. The apparent relationship involving
the Seyfert\,2 objects shows again when comparing the soft X-rays with
the warm infrared flux whereas no correlation is seen with the cold
infrared flux. 

The intermediate Seyfert type\,1.5 galaxy NGC\,4151, also shares the
same increasing trend as the Seyfert\,2 type objects.  However, the
position in the diagrams of the four Seyfert\,1 objects in the sample,
NGC\,5548, NGC\,5033, Mrk\,335 and Mrk\,817 markedly depart from the
Seyfert\,2 trend. That is in part due to the fact that their integrated soft
X-ray emission is substantially larger by more than one order of
magnitude than that from the Seyfert\,2 objects.

On the other hand, the trends above described contrast with the pure
scatter diagram arising when the comparison is done with the
absorption-corrected hard X-ray fluxes (not shown).

\vspace{-5mm}
\section{Discussion}
\vspace{-5mm}
The [O\,IV] and [Ne\,V] coronal lines discussed in this work imply the
presence of ions  with
ionization potential of 50\,eV and 100\,eV respectively.  Thus,  
a correlation  between the coronal line emission and
the soft-Xray emission is 
somewhat expected if considered that the energy required to produce
those high ionization species is very close to the soft X-ray energies
here considered.

The soft X-ray emission shows also related to the mid-infrared emission
arising from these galaxies. In Prieto et al. (2000), the mid-infrared
emission was also found closely related to the coronal line emission
arising in the nucleus of these galaxies. Both results point to the
AGN as the dominant heating source of the circumnuclear dust.  On the
other hand, circumnuclear star-forming regions are expected to further
contribute to the heating of the dust at mid-infrared wavelengths. Indeed,
ISOCAM images in the 5--8\,$\mu$m range reveal disks and/or multiple
emitting structures in some of these objects (P\'erez-Garc\'{\i}a et
al. in preparation).  Thus, some dispersion in the relation between the
soft X-ray and the mid-infrared continuum might be expected, particularly
taking into account the large aperture size of the ISOPHOT
measurements. Note that the relationship shows however more clear when
involving the ground-base 10\,$\mu$m emission, which is measured in
apertures about or smaller than $10''$.

The few Seyfert\,1 galaxies  in the sample present a soft X-ray
emission larger by at least an order of magnitude than those measured
in  their
Seyfert\,2 counterparts. Such difference seems to be a
characteristic of Seyfert\,1 galaxies: Rush et al. (1996) reported on
 systematically larger soft X-ray luminosities in Seyfert\,1 galaxies
on the basis of a much larger sample of Seyfert galaxies. 
The authors  report also on 
 a correlation -- in luminosities --  between the 12\,$\mu$m
IRAS emission and the soft X-ray emission for all the galaxies in their
sample regardless of the Seyfert type. In  the present sample,
we find that the larger soft X-ray emission of the Seyfert\,1
objects  make them already depart from the overall trend shown by the
Seyfert\,2s; subjected to the small number of Seyfert\,1 
objects considered, the
soft X-ray emission of these objects  shows indeed no trend at all with
their either  coronal line or mid-infrared emission.   
The discrepancy  is somewhat unexpected, in particular if
considered  the positive trends --in fluxes-- found within
 the Seyfert\,2 class.  If confirmed on a much larger sample, it would
indicate the presence of an additional X-ray component that is
dominant and rather anisotropic in Seyfert\,1s but unseen in
Seyfert\,2 objects.

Moving to the hard X-rays, not apparent trend  between the
absorption-corrected  emission and either
 the coronal line or the infrared continuum
emissions is detected.  
This  is somewhat expected if considered that first, 
heating of the dust from energies beyond 2\,keV would lead to hotter
dust  emitting at higher infrared frequencies; 
second, the energies required to produce the high ionization species 
considered in this work, $O^{2+}$ and $Ne^{3+}$, is well below 2\,keV.

 Contini et al. (1998 and
references therein) have modeled the
 spectral energy distribution of Seyfert\,2 galaxies and 
  shown that the re-emission
 by dust in the infrared is closely related to bremsstrahlung from hot gas
(X-ray)  in the
 vicinity of the nucleus. Indeed, the temperature of the gas and that
 of the grains are found to be closely related. In particular, in
 models that account for the combined effect of photoionization and
 shocks, soft X-ray are found mainly emitted from hot gas in the
 immediate post shock region. Mutual heating between dust and gas
 leads to the corresponding emission by dust in the near--mid infrared. The
 observed relationship between the mid-infrared  and the soft X-ray emissions
revealed by this sample   provides support to that scenario.

The detailed analysis of this work will appear in Prieto, P\'erez-Garc\'{\i}a
\& Rodr\'{\i}guez-Espinosa (2000).

\label{}

\end{document}